\documentstyle[aps,preprint,pra]{revtex}
\begin{document}
\newcommand{\etal}{{\em et al.}\/}
\newcommand{\IP}{inner polarization}
\newcommand{\IPF}{\IP\ function}
\newcommand{\IPFs}{\IP\ functions}
\newcommand{\auth}[2]{#1 #2, }
\newcommand{\twoauth}[4]{#1 #2 and #3 #4, }
\newcommand{\andauth}[2]{and #1 #2, }
\newcommand{\jcite}[4]{#1 #2 (#4) #3}
\newcommand{\et}{ and }
\newcommand{\erratum}[3]{\jcite{erratum}{#1}{#2}{#3}}
\newcommand{\JCP}[3]{\jcite{J. Chem. Phys.}{#1}{#2}{#3}}
\newcommand{\jms}[3]{\jcite{J. Mol. Spectrosc.}{#1}{#2}{#3}}
\newcommand{\jmsp}[3]{\jcite{J. Mol. Spectrosc.}{#1}{#2}{#3}}
\newcommand{\theochem}[3]{\jcite{J. Mol. Struct. ({\sc theochem})}{#1}{#2}{#3}}
\newcommand{\jmstr}[3]{\jcite{J. Mol. Struct.}{#1}{#2}{#3}}
\newcommand{\cpl}[3]{\jcite{Chem. Phys. Lett.}{#1}{#2}{#3}}
\newcommand{\cp}[3]{\jcite{Chem. Phys.}{#1}{#2}{#3}}
\newcommand{\pr}[3]{\jcite{Phys. Rev.}{#1}{#2}{#3}}
\newcommand{\jpc}[3]{\jcite{J. Phys. Chem.}{#1}{#2}{#3}}
\newcommand{\jpca}[3]{\jcite{J. Phys. Chem. A}{#1}{#2}{#3}}
\newcommand{\jcc}[3]{\jcite{J. Comput. Chem.}{#1}{#2}{#3}}
\newcommand{\molphys}[3]{\jcite{Mol. Phys.}{#1}{#2}{#3}}
\newcommand{\physrev}[3]{\jcite{Phys. Rev.}{#1}{#2}{#3}}
\newcommand{\mph}[3]{\jcite{Mol. Phys.}{#1}{#2}{#3}}
\newcommand{\cpc}[3]{\jcite{Comput. Phys. Commun.}{#1}{#2}{#3}}
\newcommand{\jcsfii}[3]{\jcite{J. Chem. Soc. Faraday Trans. II}{#1}{#2}{#3}}
\newcommand{\jacs}[3]{\jcite{J. Am. Chem. Soc.}{#1}{#2}{#3}}
\newcommand{\ijqcs}[3]{\jcite{Int. J. Quantum Chem. Symp.}{#1}{#2}{#3}}
\newcommand{\ijqc}[3]{\jcite{Int. J. Quantum Chem.}{#1}{#2}{#3}}
\newcommand{\spa}[3]{\jcite{Spectrochim. Acta A}{#1}{#2}{#3}}
\newcommand{\tca}[3]{\jcite{Theor. Chem. Acc.}{#1}{#2}{#3}}
\newcommand{\tcaold}[3]{\jcite{Theor. Chim. Acta}{#1}{#2}{#3}}
\newcommand{\jpcrd}[3]{\jcite{J. Phys. Chem. Ref. Data}{#1}{#2}{#3}}
\newcommand{\APJ}[3]{\jcite{Astrophys. J.}{#1}{#2}{#3}}
\newcommand{\astast}[3]{\jcite{Astron. Astrophys.}{#1}{#2}{#3}}
\newcommand{\arpc}[3]{\jcite{Ann. Rev. Phys. Chem.}{#1}{#2}{#3}}


\draft
\title{The ground-state spectroscopic constants of Be$_2$ revisited}
\author{Jan M.L. Martin}
\address{Department of Organic Chemistry,
Kimmelman Building, Room 262,
Weizmann Institute of Science,
76100 Re\d{h}ovot, Israel. {\em Email:} \verb|comartin@wicc.weizmann.ac.il|
}
\date{Submitted to {\it Chem. Phys. Lett.} September 17, 1998}
\maketitle
\begin{abstract}
Extensive ab initio calibration calculations combined with extrapolations
towards the infinite-basis limit lead to a ground-state dissociation energy
of Be$_2$, $D_e$=944$\pm$25 cm$^{-1}$, substantially higher than the 
accepted experimental value, and confirming recent theoretical findings.
Our best computed spectroscopic observables (expt. values in parameters)
are $G(1)-G(0)$=223.7 (223.8), $G(2)-G(1)$=173.8 (169$\pm$3),
$G(3)-G(2)$=125.4 (122$\pm$3), and $B_0$=0.6086 (0.609) cm$^{-1}$;
revised spectroscopic constants are proposed.
Multireference calculations based on a full valence CAS(4/8) 
reference space suffer
from an unbalanced description of angular correlation; for the utmost
accuracy, the $(3s,3p)$ orbitals should be added to the reference space.
The quality
of computed coupled cluster results depends crucially on the description of
connected triple excitations; the CC5SD(T) method yields unusually good
results because of an error compensation.
\end{abstract}

\section{Introduction}

Despite the small size of the beryllium dimer, Be$_2$, a correct
computational description of its 
$X~^{1}\Sigma^{+}$ ground state has long been considered
as one of the most challenging problems in quantum chemistry.\cite{Roe96}
Intuitively one would expect a purely repulsive potential between 
two closed-shell singlet atoms --- or perhaps a shallow van der Waals-like
minimum --- and in fact the Hartree-Fock potential is purely repulsive. 
However, the small $(2s)-(2p)$ gap in atomic beryllium complicates
the picture, and when angular correlation is admitted, a tightly
bound molecule is in fact found due to an avoided crossing between
$(2s)^{2}+(2s)^{2}$ and $(2s)^1(2p_z)^1+(2s)^1(2p_z)^1$ curves. As a 
result, the wave function is strongly biconfigurational, and in fact
an active space of at least four orbitals (the abovementioned plus
$(2s)^1(2p_z)^1+(2s)^{2}$ and $(2s)^{2}+(2s)^{1}(2p_z)^{1}$) is
required to obtain a qualitatively correct potential curve \cite{Sza96b}.

The Hartree-Fock limit potential is purely repulsive, and early
coupled cluster with all double excitations (CCD) 
calculations\cite{Bar78} found only a shallow van der Waals-like
minimum. Multireference configuration interaction 
studies\cite{Liu80,Blo80} on the other hand predicted a tightly
bound minimum, as did (with a highly exaggerated binding energy) a 
pioneering density functional study\cite{Jon79}. These conclusions
were corroborated in 1983 by a valence FCI (full configuration interaction)
study\cite{Har83}, and in the next year, Bondybey and English\cite{Bon84}
reported the first experimental observation. Bondybey\cite{Bon84b}
subsequently reported  
$R_e$=2.45 \AA\ and the first four vibrational quanta 
223.2, 169.7, 122.5, and 
79 cm$^{-1}$; assuming a Morse potential, he suggested a dissociation
energy of 790$\pm$30 cm$^{-1}$. Petersson and Shirley (PS)\cite{Pet89}, 
following ab initio calculations of their own, re-analyzed the 
experimental data in terms of a Morse+$1/R^6$ potential and 
suggested  an upward revision to $D_e$=839$\pm$10 cm$^{-1}$. Recent 
high-level calculations suggest even higher binding energies: for 
instance, St\"arck and Meyer\cite{Sta96} (SM), using MRCI (multireference 
configuration interaction) and a core polarization potential (CPP) 
found $D_e$=893 cm$^{-1}$ as well as $r_e$=2.448$_5$ \AA, while
MR-AQCC (multireference averaged quadratic coupled cluster\cite{aqcc})
calculations by F\"usti-Moln\'ar and Szalay \cite{Sza96} (FS) 
established $D_e$=864 cm$^{-1}$ as a lower bound. R{\o}eggen and 
Alml\"of (RA)\cite{Roe96} carried out extensive calibration 
calculations with an extended geminal model and gave 841$\pm$18 
cm$^{-1}$ as their best estimated binding energy. Evangelisti et al.
(EBG)\cite{Eva95}
carried out valence-only FCI calculations in a $[6s5p3d2f1g]$
basis set, and concluded that inner-shell correlation must contribute
substantially to the binding energy since their value (an exact 
valence-only solution within this large basis set) was still 
appreciably removed from experiment. This conclusion was confirmed by 
an all-electron FCI in a small $[9s2p1d]$ basis set (which still 
involved in excess of 10$^{9}$ determinants)\cite{Eva96}.

Part of the uncertainty in the best theoretical values resides in the 
fact that the basis sets used, while quite large, are still finite.
Convergence of angular correlation is known to be excruciatingly slow, with
an asymptotic expansion in terms of the maximum angular momentum $l$ that 
starts at $l^{-4}$ for contributions of individual angular momenta and
at $l^{-3}$ for overall $l$-truncation error\cite{Kut92}. Recently
$l$-extrapolations have been proposed\cite{l4,c2h4tae} which permitted
the calculation of total atomization energies of small polyatomic 
molecules with mean absolute errors as low as 0.12 kcal/mol. Among 
other applications, this method made possible a definitive 
re-evaluation\cite{bf3} of the heat of vaporization of boron from a calibration
quality calculation on BF$_3$. 

In the present work, we apply this method to the dissociation energy 
of Be$_2$. It will be shown that the valence-only basis set limit
is in fact as large as 875$\pm$10 cm$^{-1}$, and the overall $D_e$ as
large as 945$\pm$20 cm$^{-1}$. 

\section{Methods}

The multireference and FCI calculations, as well as those
using the CCSD(T)\cite{Rag89} coupled cluster method, were
carried out 
using a prerelease version of MOLPRO97 
\footnote{
MOLPRO 97.3 is a package of {\em ab initio} programs written by
\auth{H.-J.}{Werner} \andauth{P. J.}{Knowles}
with contributions from
\auth{J.}{Alml\"of} \auth{R. D.}{Amos} \auth{A.}{Berning} \auth{D. L.}{Cooper}
\auth{M. J. O.}{Deegan} \auth{A. J.}{Dobbyn}
\auth{F.}{Eckert} \auth{S. T.}{Elbert} \auth{C.}{Hampel} \auth{R.}{Lindh}
\auth{A. W.}{Lloyd} \auth{W.}{Meyer} \auth{A.}{Nicklass}
\auth{K. A.}{Peterson} \auth{R. M.}{Pitzer} \auth{A. J.}{Stone}
\auth{P. R.}{Taylor} \auth{M. E.}{Mura} \auth{P.}{Pulay}
\auth{M.}{Sch\"utz} \auth{H.}{Stoll} and \auth{T.}{Thorsteinsson}} 
running on an SGI Origin 2000 minisupercomputer at the Weizmann 
Institute of Science. Calculations with other coupled cluster
methods were carried out using 
ACES II\footnote{
\auth{J. F.}{Stanton} \auth{J.}{Gauss}
\auth{J. D.}{Watts} \auth{W.}{Lauderdale} \andauth{R. J.}{Bartlett}
(1996) ACES II, an ab initio program system, incorporating the
MOLECULE vectorized molecular integral program by J. Alml\"of, J. and P. R. Taylor,
and a modified version of the ABACUS integral derivative package by
T. Helgaker, H. J. Aa. Jensen, P. J{\o}rgensen, J. Olsen, and P. R. Taylor.
} running on a DEC Alpha workstation.

Most basis sets used belong to the correlation consistent polarized 
valence $n$-tuple zeta (cc-pV$n$Z) family of Dunning\cite{Dun89}.
The cc-pVDZ, cc-pVTZ, cc-pVQZ and cc-pV5Z basis sets are $[3s2p1d]$, $[4s3p2d1f]$,
$[5s4p3d2f1g]$, and $[6s5p4d3f2g1h]$ contractions, respectively, of
$(9s4p1d)$, $(11s5p2s1d)$, $(12s6p3d2f1g)$, and $(14s8p4d3f2g1h)$
primitive sets. For assessing inner-shell correlation effects, we
used the core correlation basis set of Martin and Taylor\cite{hf}:
MTvtz and MTvqz denote completely uncontracted cc-pVTZ and cc-pVQZ basis sets,
respectively, augmented with one tight $p$, three tight $d$, and two 
tight $f$ functions with exponents derived by successively 
multiplying the highest exponent already in the basis set with a 
factor of three. The MTv5z basis set is obtained similarly, but in 
addition has a single tight $g$ function as well.

\section{Results and discussion}

\subsection{Valence electron contribution}

For the cc-pVDZ, cc-pVTZ, and cc-pVQZ basis sets, valence-only FCI calculations 
could be carried out. The results at the reference geometry $R=2.45$ 
\AA\ are given in Table 1.

By comparison with CCD, CCSD\cite{Pur82}, and CCSDT\cite{ccsdt} 
results in the same basis sets
(CCSDTQ being equivalent to FCI for this case), we can partition the 
valence binding energy into contributions from connected single, 
double, triple, and quadruple excitations as well as investigate their 
basis set convergence. As previously noted by Sosa et al.\cite{Sos88}
in small basis sets, no covalent binding is seen at the CCSD level; 
they found CCSDT-1\{a,b\} and CCSDT-2 to display only a shallow 
ripple, while CCSDT-4 slightly exaggerates the potential well and 
full CCSDT is slightly above the FCI result. These conclusions
are confirmed here; moreover, as the basis set is increased, the
CCSDT results closely track the FCI ones, which in this case implies
that the contribution of connected quadruples to the binding converges
very rapidly to an estimated basis set limit of 85 cm$^{-1}$. By 
contrast, the contribution of connected triples is actually 
substantially larger than the atomization energy itself, and is 
apparently not yet converged with the cc-pVQZ basis set. 

Our attempts to carry out a CCSDT/cc-pV5Z calculation with the available
computer infrastructure met with failure. CCSD(T) calculations are an 
obvious alternative, but are seen in Table 1 to on the one hand 
underestimate the importance of connected triple excitations, and on 
the other hand to display considerable basis set dependence in the
difference with full CCSDT (hence making it a poor candidate for 
extrapolation). The difference between CCSD(T) and CCSDT starts at
fifth order in perturbation theory; in the method alternatively
known as CCSD+T(CCSD)*\cite{ccsd+tq} 
and, in Bartlett's recent notation\cite{Bar95},
CC5SD(T), the missing $E_{5TT}$ term is included quasiperturbatively
at a computational
expense scaling as $n_{\rm occ}^{3}n_{\rm virt}^{5}$. As seen in Table 
1, CC5SD(T) slightly overestimates the connected triple excitations 
contribution but does so in a highly systematic manner, the difference
being constant between 38 and 40 cm$^{-1}$. Because of an error 
compensation with neglect of connected quadruple excitations, it is 
actually the one single-reference method short of full CI that we
find to be closest to the exact solution. In short, it is the
ideal candidate for basis set extrapolation. 

The CCSD+TQ(CCSD)* or CC5SD(TQ) method, which includes the leading 
contribution of connected quadruple excitations in a 
similar fashion, appears to seriously overestimate it, and we have 
not considered it further. 

Basis set superposition error for the valence electrons was considered
using the standard counterpoise (CP) correction\cite{Boy70}. In the 
present case, it drops from 36 cm$^{-1}$ (cc-pVDZ) over 24 (cc-pVTZ) to 6 
cm$^{-1}$ for the cc-pVQZ basis set, and a paltry 3.5 cm$^{-1}$ for the cc-pV5Z
basis set. 

From the FCI/cc-pV\{D,T,Q\}Z results, we may attempt extrapolation, 
either from the uncorrected $D_e$ values (assuming that the extrapolation
will absorb BSSE which strictly vanishes at the basis set limit) or
after subtracting the counterpoise correction in each case. With a
variable-$\alpha$ 3-parameter correction, this leads to basis set
limits of 841 and 859 cm$^{-1}$, respectively. Using the simple
$A+B/l^{3}$ formula\cite{Hal98} on just the final two results, we
obtain values of 863 (raw) and 870 (CP-corrected) cm$^{-1}$. 

It can rightly be argued that the cc-pVDZ basis set is really too small to 
be involved in this type of extrapolation, and that a cc-pV5Z result is 
essential for this purpose. This requires us to estimate an
FCI/cc-pV5Z result from the additivity 
approximation Method/cc-pV5Z$+$FCI/cc-pVQZ$-$Method/cc-pVQZ. With Method=CC5SD(T),
we obtain $D_e$(FCI/cc-pV5Z)$\approx$818.2 cm$^{-1}$; 3-point extrapolation
yields 881 cm$^{-1}$ for the raw, and 872 cm$^{-1}$ for the 
CP-corrected, results as the basis set limit. Using the simple
$A+B/l^{3}$ formula, we obtain the alternative results 857 and 873 
cm$^{-1}$, respectively. The fact that the two extrapolations yield 
essentially the same result for the CP-corrected values, as well as 
that they are in very close agreement with the results with the 
smaller basis sets, is very satisfying.

It could likewise be argued that in fact the SCF and correlation 
contributions should be handled separately\cite{c2h4tae}, with an
exponential or $(l+1/2)^{-5}$ formula for the SCF contribution and
an $A+B/(l+1/2)^{\alpha}$ or $A+B/l^{3}$ formula for the correlation
contribution alone. We then find that the SCF contribution, with the
cc-pV5Z basis set, lies within 3 cm$^{-1}$ of the numerical HF limit;
after adding in the basis set limits for the correlation contribution,
we obtain, after counterpoise correction, 869 cm$^{-1}$ with the
3-point and 871 cm$^{-1}$ with the 2-point formula. 

One further objection would be to the use of even a high-level 
single-reference method for a problem that is intrinsically 
multireference in character. We have therefore considered MRCI (multireference
configuration interaction) augmented with the multireference Davidson
correction\cite{Blo83}, MRACPF\cite{Gda88} (multireference averaged
coupled pair functional), and MRAQCC\cite{aqcc} (multireference averaged
quadruples coupled cluster) methods with a variety of active spaces.
A 4/4 active space appears to be unsatisfactory for our purposes;
hence we have considered full-valence 
CAS(4/8)-ACPF (averaged coupled pair functional\cite{Gda88}) and
CAS(4/8)-AQCC as alternatives. Except for the cc-pVDZ basis set, both 
methods seem to track the FCI results quite closely, with CAS(4/8)-ACPF 
accidentally coinciding with the FCI results. Again applying the
same additivity approximation as above, we obtain estimated FCI/cc-pV5Z
results from these calculations of 821.5 and 819.6 cm$^{-1}$, 
especially
the latter quite close to the CC5SD(T) derived value. 

Interestingly, the CAS(4/8)-ACPF wave function contains a fairly large
number of external excitations with fairly high amplitudes, most of
them involving excitation into (3p)-type Rydberg orbitals. Inspection
of the atomic wave function for Be atom revealed that excitations
into the fairly low-lying (3p) orbitals have amplitudes as large as
0.09 (for each of three symmetry-equivalent components); since in addition
the (3s) orbital is below the (3p) orbital in energy and there appears to
be no clear separation between $(3s)$- and $(3p_z)$-derived $\sigma$
orbitals, this suggests a (4/16) active space which spans all
molecular orbitals derived from atomic $(2s,2p,3s,3p)$ orbitals.
External excitations now carry so little weight in the wave function
that CAS(4/16)-MRCI+Dav, CAS(4/16)-ACPF and CAS(4/16)-AQCC yield
essentially identical results. Arbitrarily selecting the 
CAS(4/16)-ACPF result for extrapolation, we obtain a best estimate
of 821.5 cm$^{-1}$ for the FCI/cc-pV5Z $D_e$. After counterpoise correction, the
CAS(4/16)-ACPF derived value leads to a basis set limit value of
885.6 cm$^{-1}$ with the 3-point and 861.4 cm$^{-1}$ with the 2-point
formula. Taking the average of the latter two values and the CC5SD(T)
derived ones, we finally propose 872$\pm$15 cm$^{-1}$ as our best
estimate for the valence-only $D_e$.

As a final remark, let it be noted that the extrapolations in all cases
bridge an area of no more than 50--70 cm$^{-1}$; by substituting 
$l=6$ in the extrapolation fomulas, we can estimate that
calculations with the
next large basis set, cc-pV6Z (i.e. [7s6p5d4f3g2h1i]), 
would only recover about 20--25 cm$^{-1}$ of that total.

\subsection{Inner-shell contribution}

By taking the difference between their computed MRCI results with and
without the core polarization potential, SM found that inner-shell 
correlation would add 0.38 m$E_h$, or 83 cm$^{-1}$, to the 
atomization energy. RA computed a contribution of $(1s)$ correlation
(almost exclusively core-valence correlation) of 0.40654 m$E_h$, or
89.2 cm$^{-1}$. 

Our results for the effect of inner-shell correlation are collected in
Table 2.
Using the MTvtz, MTvqz, and MTv5z basis sets in succession at the 
CAS(4/16)-ACPF level.
we find contributions of inner-shell correlation
to the binding energy of 82.1, 80.6, and 77.8 cm$^{-1}$. BSSE 
contributions to the core-correlation contribution (taken as the
difference between all-electron and valence-only BSSEs in the same
basis set) are 3.8, 2.9, and 1.5 cm$^{-1}$, respectively, such
that the counterpoise-corrected values of 78.3, 77.7, and 76.3 cm$^{-1}$
appear to be quite handsomely converged. 

For comparison, the counterpoise-corrected CCSD(T) results are 
75.0, 73.1, and 70.9 cm$^{-1}$, while a 
CC5SD(T)/MTvtz calculation yielded 63.3 cm$^{-1}$ without 
counterpoise correction. CAS(4/4)-ACPF and CAS(4/8)-ACPF calculation
actually yielded small {\em negative} inner-shell correlation 
contributions which are clearly an artifact of the reference space.

We also note that the counterpoise-corrected all-electron
CAS(4/16)-ACPF/cc-pV5Z $D_e$ of 882.4 kcal/mol is already higher
than the FS number, and in fact near the SM value. Indeed, since 
this level of electron correlation appears to systematically underestimate
the valence binding energy by 15--16 cm$^{-1}$ compared to FCI (see Table 1),
we can establish 900 cm$^{-1}$ as a lower limit to $D_e$.

Adding the best inner-shell correlation energy contribution of 76.2 cm$^{-1}$
to our best valence binding energy, we obtain a best estimate for the
all-electron binding energy of 948$\pm$20 cm$^{-1}$, where the increased
error bar reflects the added uncertainty in the inner-shell contribution.

The effect of scalar relativistic effects was gauged from Darwin and 
mass-velocity terms obtained from CAS(4/16)-ACPF/MTvqz calculations
by perturbation theory\cite{Mar83}. At $-$4.0 cm$^{-1}$, it is essentially
negligible.

Combining our best estimates for valence, inner-shell, and 
relativistic contributions, we finally obtain a best estimate for
$D_{e}$(Be$_{2}$) of 944 $\pm$25 cm$^{-1}$, which suggests that 
the PS value for $D_e$ may need to be revised upward by as much as 100 
cm$^{-1}$.

\subsection{Potential curve}

Computed bond distances $r_e$, harmonic frequencies $\omega_e$, and
the first three anharmonicities $\omega_ex_e$, $\omega_ey_e$, and
$\omega_ez_e$ are collected in Table 3. They were obtained by a 
Dunham analysis on eighth-order polynomials fitted to some 25 
computed energies at bond distances
spaced around the putative minimum with distances of 0.02 \AA.

While good fits could be obtained to the CCSD(T) and CC5SD(T)
results, attempts to fit CAS(4/8)-\{MRCI,ACPF,AQCC\} curves
in the same manner met with failure. No such problem was
encountered with results based on a smaller CAS(4/4) reference
wave function: investigation of the CASSCF energies revealed that
while the CAS(4/4) curve is bound, the CAS(4/8) curve is purely
repulsive in the region sampled. Further investigation revealed 
that with increasing $r$, amplitudes for excitations into $(3p)$
derived Rydberg orbitals progressive take on pathological 
dimensions (as large as 0.35): under such circumstances, the
noisy character of the CAS(4/8)-ACPF potential curves should
not come as a surprise. As expected, expanding the reference space
to CAS(4/16) eliminates the problem, as well as restores
a bound CASSCF potential curve. Apparently the (2p) and (3p)
orbitals are close enough in importance that a balanced reference
space requires that they either be both included or both excluded.

From comparing CAS(4/16)-ACPF/cc-pVTZ and FCI/cc-pVTZ spectroscopic
constants, it is obvious that the former treatment is indeed very
close to an exact solution and the method of choice for 1-particle
basis set calibration. CC5SD(T) yields surprisingly good $r_e$ and
$\omega_e$ values (in fact agreeing more closely with FCI than CCSDT)
but strongly overestimates the anharmonicity of
the curve. Performance of CCSD(T) is fairly poor, although the quality
of the results is still amazing considering the pathological character
of the molecule.

Extension of the basis set to cc-pVQZ has a very significant effect
on the spectroscopic constants, with $r_e$ being shortened by 0.026 \AA\ 
and $\omega_e$ going up by 16 cm$^{-1}$. Further extension to cc-pV5Z
has a much milder effect, and suggests that convergence is being approached
for the molecular properties. $A+B/l^3$ extrapolation suggests that further
basis set extension may affect $r_e$ by a further $-$0.003 \AA\ and increase
$\omega_e$ by another $+$2 cm$^{-1}$. 

Ideally, we would have liked to present all-electron CAS(4/16)-ACPF/MTv5z
curves in order to include inner-shell correlation. Since however a single 
point in such a curve took more than a day of CPU time on an SGI Origin 2000,
we have not pursued this option further, and have instead contented ourselves
with considering the difference between CCSD(T)/MTv5z curves with and without
constraining the $(1s)$-like orbitals to be doubly occupied. Our results 
suggest that inner-shell correlation reduces $r_e$ by 0.03 \AA\ and increases
$\omega_e$ by 14 cm$^{-1}$. The spectroscopic constants given as `best
estimate' are obtained by adding these contributions to the extrapolated
CAS(4/16)-ACPF/cc-pV$\infty$Z
results, as well as the small difference between FCI/cc-pVTZ and 
CAS(4/16)-ACPF/cc-pVTZ.

Obviously, given the highly anharmonic nature of the potential surface,
a Dunham-type perturbation theory analysis is not appropriate. Like in
our recent calibration study on the first-row diatomic hydrides, we have
transformed our 8th-order Dunham expansion and computed dissociation energy
to a variable-beta Morse (VBM) potential\cite{Cox92} 
\begin{equation}
V_c = D_e \left(1-\exp[-z (1 + b_1 z + b_2 z^2 + \ldots+b_6 z^6)]\right)^2
\end{equation}
in which $z\equiv \beta (r-r_e)/r_e$ and the parameters $b_n$ and $\beta$
are obtained by derivative matching as discussed in detail in Ref.\cite{ch}.
The one-dimensional Schr\"odinger equation was then integrated using 
the algorithm of Balint-Kurti et al.\cite{balint}, on a grid of
256 points over the interval $[0.2r_e,3r_e]$. 

The results for the first four vibrational quanta are given in Table \ref{level}. 
We have considered three potentials. The first two are the uncorrected FCI/cc-pVTZ and
CAS(4/16)-ACPF/cc-pV5Z potentials; the third one was obtained by 
substituting our best estimate $D_e$
and $r_e$, and adjusting $\beta$ such that the best estimate $\omega_e$
is matched. (The $b_n$ remain unchanged from the CAS(4/16)-ACPF/cc-pV5Z values.) What this
latter approaches in effect assumes is that the {\em shape} of the CAS(4/16)-ACPF/cc-pV5Z
curve is fundamentally sound.

As expected, the unadjusted FCI/cc-pVTZ potential seriously underestimates 
the first three vibrational quanta because of the strong dependence of
$D_e$, $\omega_e$, and $r_e$ on the basis set and the inclusion of inner-shell
correlation. CAS(4/16)-ACPF/cc-pV5Z does so to a much lesser extent.
Our `best estimate' potential, however, reproduces the fundamental (the only
transition known with some precision) essentially exactly, and is in
good agreement with experiment for the next two quanta. Since the VBM form
of the potential does not take into account long-distance behavior and
the fourth quantum lies at 80\% of the dissociation energy, it is not
surprising that the fourth quantum is seriously overestimated. 

Finally, let us turn to the spectroscopic constants derived from our
best potential (Table 5).
Our best $\omega_e$ is in perfect agreement with SM but substantially 
lower than the Bondybey value. Our best $\omega_ex_e$ is substantially
smaller than both the Bondybey and SM values: however, both of the 
latter were determined phenomenologically as 
$[G(2)-2G(1)-G(0)]/2$ and therefore include 
contributions from higher-order anharmonicities. If we compute the 
same quantity, we obtain perfect agreement with the SM value. 
While our rotation-vibration coupling constant $\alpha_e$ is in
very good agreement with the SM calculations, it is substantially
larger than the Bondybey value. However, it should be noted that
the Be$_2$ potential is so anharmonic that the series
$B_n=B_e-\alpha_e(n+1/2)+\gamma_e(n+1/2)^2+\delta_e(n+1/2)^3+\ldots$
cannot be truncated after the linear term; from our best computed
spectroscopic constants, we obtain $B_0$=0.6086 cm$^{-1}$, in 
perfect agreement with Bondybey's value of 0.609 cm$^{-1}$ for this
{\em observable} quantity. In short,
we argue that our computed $r_e=2.440$ \AA\
is more reliable than the Bondybey value of 2.45$_0$ \AA.

As a final note, we point out that this revised reference geometry 
($r_{e}$=2.440 \AA) would
not have affected our calculation of $D_e$ materially, since the
energy difference between $R=$2.44 and $R=2.45$ \AA\ with our best
potential only amounts to 0.4 cm$^{-1}$.

\section{Conclusions}

From an exhaustive basis set convergence study on the dissociation
energy of the ground-state Be$_2$, we find that the accepted 
experimental value needs to be revised upward to a best estimate
of 944 $\pm$25 cm$^{-1}$. Individual contributions to this value
include a valence-only FCI basis set limit of 872$\pm$15 cm$^{-1}$,
an inner-shell contribution of 76$\pm$10 cm$^{-1}$, and relativistic
corrections as small as $-$4  cm$^{-1}$. The 
performance of single-reference methods for this molecule is crucially
dependent on their treatment of connected triple excitations; while 
CCSD(T) underestimates binding in this molecule, the CC5SD(T) method
performs surprisingly well at a fraction of the cost of full CCSDT.
The contribution of connected quadruple excitations is small (80 
cm$^{-1}$) and fairly insensitive to the basis set. 
Accurate multireference calculations require an active space
which treats angular (2p,3p) correlation in a balanced way; a full-valence
CAS(4/8) reference does {\em not} satisfy this criterion.
For the utmost
accuracy, a CAS(4/16) reference including the $(3s,3p)$ orbitals is required,
while for less accurate work a CAS(4/4) reference is recommended. 
Our best computed spectroscopic observables (expt. values in parameters)
are $G(1)-G(0)$=223.7 (223.8), $G(2)-G(1)$=173.8 (169$\pm$3),
$G(3)-G(2)$=125.4 (122$\pm$3), and $B_0$=0.6086 (0.609) cm$^{-1}$.
Our best computed spectroscopic constants represent substantial
revisions from the experimentally derived values; in particular, the 
bond length is 0.01 \AA\ shorter than the accepted experimental
value.

\acknowledgments

The author is a Yigal Allon Fellow, the incumbent of the Helen and Milton
A. Kimmelman Career Development Chair, and
an Honorary Research Associate (``Onderzoeksleider
in eremandaat'') of the
National Science Foundation of Belgium (NFWO/FNRS). 
He acknowledges support from the Minerva Foundation, Munich, Germany.
This study was inspired by discussions with 
Dr. Russell D. Johnson III (NIST) on the poor 
performance of standard computational thermochemistry 
methods.


\begin{table}
\caption{Convergence of the valence dissociation energy 
(cm$^{-1}$) of Be$_2$ as a function of
basis set and electron correlation treatment}
\squeezetable
\begin{tabular}{lrrrrr}
 & cc-pVDZ & cc-pVTZ & cc-pVQZ & cc-pV5Z \\
FCI & 23.92 & 630.50 & 764.81 & ---\\
\hline
 & \multicolumn{3}{c}{\underline{~~~~~~~~~~~~Difference with FCI~~~~~~~~~~~~~}} & Actual & Estimated \\
 &  & &  & value & FCI/cc-pV5Z$^a$ 
 \\
\hline
SCF & -2759.54 & -3277.17 & -3396.48 & -2626.13 & 770.36 \\
CAS(4/8)-CI+Davidson & 38.85 & -36.27 & -49.38 & 769.23 & 818.61 \\
CAS(4/4)-ACPF & 35.66 & -56.63 & -69.67 & 747.16 & 816.82 \\
CAS(4/4)-AQCC & 22.10 & -84.61 & -98.23 & 717.18 & 815.41 \\
CAS(4/8)-CI+Davidson & 84.77 & 35.53 & 36.96 & 859.67 & 822.71 \\
CAS(4/8)-ACPF & 60.02 & 0.22 & -0.50 & 821.03 & 821.53 \\
CAS(4/8)-AQCC & 43.41 & -23.76 & -24.71 & 794.89 & 819.60 \\
CAS(4/16)-CI+Davidson & 48.12 & -14.17 & -14.16 & 807.49 & 821.65 \\
CAS(4/16)-ACPF & 47.94 & -14.78 & -15.05 & 806.47 & 821.52 \\
CAS(4/16)-AQCC & 48.26 & -15.25 & -15.61 & 805.80 & 821.40 \\
CCD & -978.97 & -1245.24 & -1278.06 & -474.45 & 803.61 \\
CCSD & -943.86 & -1118.72 & -1158.37 & -353.16 & 805.21 \\
CCSDT & -73.43 & -83.13 & -84.47 & --- & --- \\
CCSD(T) & -261.65 & -257.57 & -241.37 & 587.41 & 828.78 \\
CC5SD(T) & -39.75 & -38.15 & -40.07 & 778.09 & 818.16 \\
\hline
BSSE$^b$ & 36.00 & 24.37 & 6.10 & 3.47 \\
\end{tabular}

(a) according to FCI/cc-pV5Z $\approx$ Method/cc-pV5Z + FCI/cc-pVQZ $-$ Method/cc-pVQZ

(b) counterpoise method

\end{table}

\begin{table}
\caption{Contribution of inner-shell correlation to the dissociation energy (cm$^{-1}$)
of Be$_2$}
\squeezetable
\begin{tabular}{llrrr}
Method         & $e^-$ correlated & MTvtz & MTvqz & MTv5z \\
\hline
CCSD(T)        & all        & 507.36 & 614.26 & 661.56 \\
               & valence    & 432.34 & 541.15 & 590.68 \\      
               & difference &  75.03 &  73.11 &  70.88 \\
CC5SD(T)       & all        & 641.77 \\
               & valence    & 705.06 \\
               & difference &  63.29 \\
CAS(4/4)-ACPF  & all        & 580.39 & 673.94 &        \\
               & valence    & 588.89 & 676.80 &        \\
               & difference &  -8.50 &  -2.86 &        \\
CAS(4/8)-ACPF  & all        & 679.29 & 773.33 & 823.22 \\
               & valence    & 682.17 & 779.23 & 811.44 \\
               & difference &  -2.88 &  -5.90 & -11.78 \\
CAS(4/16)-ACPF & all        & 749.67 & 845.37 & 886.56 \\
               & valence    & 667.56 & 764.81 & 808.71 \\
               & difference &  82.11 &  80.56 &  77.85 \\
BSSE (a)       & all        &   9.43 &  7.22  &   4.06 \\
               & valence    &   5.63 &  4.36  &   2.52 \\
               & difference &   3.80 &  2.86  &   1.54 \\
\end{tabular}

(a) on CAS(4/16)-ACPF values
\end{table}

\begin{table}
\caption{Convergence of computed bond distance (\AA) and vibrational
spectroscopic constants (cm$^{-1}$) of Be$_2$ as a function of 
basis set and electron correlation treatment}
\squeezetable
\begin{tabular}{llrrrrr}
Method & Basis & $r_e$ & $\omega_e$ & $\omega_ex_e$ & $\omega_ey_e$ & $\omega_ez_e$\\
\hline
CC5SD(T) & cc-pVDZ & 2.5736 & 187.4 & 33.175 & -4.937 & \\
 & cc-pVTZ & 2.5012 & 230.8 & 23.198 & -1.179 & -0.116\\
 & cc-pVQZ & 2.4745 & 245.3 & 21.825 & -1.072 & \\
 & cc-pV5Z & 2.4718 & 247.5 & 21.367 & -0.959 & \\
CCSD(T) & MTvtzALL$^a$ & 2.4829 & 229.2 & 24.689 & -1.603 &\\ 
 & MTvtzVAL$^b$ & 2.5145 & 214.3 & 25.919 & -1.871 & \\
 & difference & -0.0316 & 14.9 & -1.230 & 0.268 & \\
 & MTvqzALL & 2.4685 & 241.5 & 22.987 & -1.316 & \\
 & MTvqzVAL & 2.4986 & 227.2 & 23.905 & -1.514 & \\
 & difference & -0.0301 & 14.3 & -0.918 & 0.198 & \\
 & MTv5zALL & 2.4652 & 243.5 & 22.721 & -1.214 & \\
 & MTv5zVAL & 2.4950 & 229.5 & 23.482 & -1.335 & \\
 & difference & -0.0298 & 14.0 & -0.761 & 0.120 & \\
FCI & cc-pVDZ & 2.5598 & 193.9 & 31.174 & -4.082 & \\
 & cc-pVTZ & 2.5021 & 234.3 & 22.383 & -1.071 & -0.097 \\
CAS(4/16)-ACPF & cc-pVTZ & 2.5041 & 232.4 & 22.639 & -1.103 & -0.093 \\
 & cc-pVQZ & 2.4781 & 246.7 & 21.325 & -1.011 & -0.084 \\
 & cc-pV5Z & 2.4750 & 249.1 & 20.856 & -0.905 & -0.061 \\
 & cc-pV$\infty$Z (c) & 2.4718 & 251.7 & 20.365 & -0.793 & -0.037 \\
Best estimate (d) & 2.4397 & 267.9 & 19.191 & -0.563 & -0.042 \\
Bondybey\cite{Bon84} & 2.45$_0$ & 275.8 & 26.0 \\
SM\cite{Sta96} & 2.448$_5$ & 268.2 & 24.9 & \\
\end{tabular}

(a) all electrons correlated

(b) only valence electrons correlated

(c) extrapolated according to $A+B/l^{3}$

(d) CAS(4/16)-ACPF/cc-pV$\infty$Z+[FCI/cc-pVTZ$-$CAS(4/16)-ACPF/cc-pVTZ]
+[CCSD(T)/MTv5zALL$-$CCSD(T)/MTv5zVAL]
\end{table}

\begin{table}
\caption{Computed and observed vibrational energy level differences (cm$^{-1}$) for the
$X~^1\Sigma^+$ state of Be$_2$\label{level}}
\begin{tabular}{lcccccccc}
     &FCI/ & CAS(4/16)-ACPF/ &best  &Expt. & \cite{Sza96}&  \cite{Pet89}&\cite{Sta96} &\cite{Roe96}\\
     &cc-pVTZ&cc-pV5Z &       (a)\\
\hline
ZPE  &110.6 &118.5       &   127.9  &      & 125  &      & 124.8 & \\
$G(1)-G(0)$&185.4 &204.2       &   223.7  & 223.8& 218  &  213 & 218.4 &221.0\\
$G(2)-G(1)$&125.5 &153.5       &   173.8  & 169  & 168  &  167 & 168.6 &162.9\\
$G(3)-G(2)$& 72.2 &109.5       &   125.6  & 122  & 112  &  122 & 112.1 & 94.2\\
$G(4)-G(3)$& 75.3 & 99.2       &   106.9  &  79  &  67  &   78 &  69.4 & 54.7\\
\end{tabular}

(a) from FCI/cc-pVTZ potential in form eq.(1), but with best estimate $r_e$,
$D_e$, and $\omega_e$ substituted according to
$\beta_{\rm new}/\beta_{\rm old}=\omega_{e,{\rm new}}r_{e,{\rm new}}\sqrt{D_{e,{\rm old}}}
/\omega_{e,{\rm old}}r_{e,{\rm old}}\sqrt{D_{e,{\rm new}}}$

\end{table}

\begin{table}
\caption{Potential function parameters in eq.(1) and mechanical 
spectroscopic constants of Be$_{2}$ with this potential. All values in
cm$^{-1}$ except $\beta$ and the $b_{n}$, which are dimensionless}
\squeezetable
\begin{tabular}{lcccc}
best potential   & calculated                   & Bondybey\cite{Bon84} 
&SM\cite{Sta96}\\
\hline
$D_e$=944.0 (a)      &   $Y_{00}$=-0.788            &  \\
$r_{e}$=2.439685     &   $\omega_e$=267.93          &  275.8  & 268.2\\
$\beta$=5.499750     &$\omega_ex_e$=20.681 (d)      &    26.0 & 24.9\\
$b_1$= 0.019920      &$\omega_ey_e$=-0.827          &    ---  \\
$b_2$=-0.048391      &$\omega_ez_e$=-0.052          &    ---  \\
$b_3$=-0.016734      &$B_e$=0.62853                 &   0.623 & 0.6213\\
$b_4$= 0.000693      &$B_0$=0.60863                 &   0.609 & \\
$b_5$= 0.001938      &$\alpha_e$=0.03787    (b)     &   0.028 & 0.037\\
$b_6$= 0.000324      &$\gamma_e$=-0.00361           &   ---   \\
                     &$\delta_e$=-0.00050           &   ---   \\
 &$D_e$=13.84$\times10^{-6}$ (c)&  14.8   \\
 &$\beta_e$=3.48$\times10^{-6}$ &  ---    \\
\end{tabular}

(a) dissociation energy

(b) $-(B_1-B_{0})$=0.02904 cm$^{-1}$

(c) quartic centrifugal distortion constant

(d) $[G(2)-2G(1)-G(0)]/2$=24.95 cm$^{-1}$

\end{table}

\end{document}